\documentclass{elsart}
\usepackage{graphicx}
\usepackage{bm}
\usepackage{psfrag}
\usepackage{amsmath}
\usepackage{amsfonts}
\usepackage{amssymb}
\usepackage[latin1]{inputenc}

\newcommand{\ket}[1]{|#1\rangle}

\newcommand{\op}[1]{|#1\rangle\langle#1|}
\newcommand{\opp}[2]{|#1\rangle\langle#2|}

\begin{document}
\begin{frontmatter}
\title{Effects of non-local initial 
conditions\\ in the Quantum Walk on the line}
\author{G. Abal\thanksref{ca}}, 
\thanks[ca]{Corresponding author: abal@fing.edu.uy}
\author{R. Donangelo\thanksref{pa}},
\author{A. Romanelli} and 
\author{R. Siri}

\thanks[pa]{Permanent address: Instituto de Física,
Universidade Federal do Rio de Janeiro, 
C.P. 68528, 21941-972 Rio de Janeiro, Brazil}

\address{Instituto de Física, Universidad de la República \\ 
C.C. 30, C.P. 11000, Montevideo, Uruguay}


\begin{abstract}
We report an enhancement of the decay rate of the survival probability when non-local initial conditions in position space are considered in the Quantum Walk on the line. It is shown how this interference effect can be understood analytically by using previously derived results. Within a restricted position subspace,  the enhanced decay is correlated with a maximum asymptotic entanglement level while the normal decay rate corresponds to initial relative phases associated to a minimum level. 
\end{abstract}

\end{frontmatter}

\section{Introduction}
\label{sec:intro}

The best known algorithms to solve ``hard"  (NP-complete) problems are based on the random walk (RW) \cite{Schoning}. This fact has motivated a strong interest in the quantum version of this process, the quantum walk (QW) \cite{Kempe03}, since it was originally proposed by Aharonov,  Davidovich and Zagury \cite{Davidovich}. The QW is a faster process than the RW \cite{Kempe02,qw-markov} and optimal quantum search algorithms have been based on it \cite{Shenvi03}. In the discrete-time QW on a line, the coin-flipping operation of the RW is replaced by a unitary operation on a coin degree of freedom, encoded in one qubit. The Hilbert space, $\protect{{\mathbf H}={\mathbf H_P}\otimes{\mathbf H_C}}$, is composed of a spatial subspace, ${\mathbf H_P}$, spanned by the orthonormal set $\{\ket{x}\}$ where the integers $x=0,\pm 1,\pm 2\ldots$ label discrete positions on the line and ${\mathbf H_C}$ is a single-qubit coin space, spanned by two orthonormal vectors $\{\ket{R}, \ket{L}\}$. A generic state for the walker is 
\begin{equation}
\ket{\Psi}=\sum_{x=-\infty}^\infty \left[ a_x\ket{R} + b_x\ket{L}\right]\otimes\ket{x}. \label{sp-wv}
\end{equation} 
One step of the quantum walker, $\ket{\Psi(t+1)}=U\ket{\Psi(t)}$, consists of a unitary operation on ${\mathbf H_C}$ followed by a conditional translation on the line,  
\begin{equation}
U=\left[\sum_{x=-\infty}^\infty
\opp{x+1}{x}\otimes\op{R}+\opp{x-1}{x}\otimes\op{L}\right]\cdot\left(I_P\otimes H\right), 
\label{Uop}
\end{equation}
where $I_P$ is the identity operator in ${\mathbf H}_P$. We particularize the unitary operation applied on the coin subspace to a Hadamard operation, $\protect{H\ket{R}=\left(\ket{R}+\ket{L} \right)/\sqrt{2}}$ and $\protect{H\ket{L}=\left(\ket{R}-\ket{L} \right)/\sqrt{2}}$. For this particular unbiased coin, the QW is known as a Hadamard walk. The evolution of an initial state $\ket{\Psi(0)}$ is given by $\ket{\Psi(t)}=U^t\ket{\Psi(0)}$,  where the non-negative integer $t$ counts the discrete time steps that have been taken. The probability distribution for finding the walker at site $x$ at time $t$ is $\protect{P(x,t)=|a_x|^2+|b_x|^2}$. The variance of this distribution increases quadratically with time as opposed to the classical random walk in one dimension, in which the increase is only linear. This advantage in the spread of the quantum walker depends strictly on quantum interference effects and is eventually lost in the presence of decoherence \cite{deco}. 

Most previous work on the subject has focused on the properties of the probability distribution $P(x,t)$ when the quantum walker is initially localized at the origin, i.e. $a_x=b_x=0$ in eq.~(\ref{sp-wv}), unless $x=0$. In this work, we consider the impact of non-local initial conditions on two different aspects of the QW: the decay of the survival probability and the level of asymptotic entanglement.

\section{Survival Probability}

The survival probability of a quantum walk in a symmetric range $[-s,s]$ of the line ($s$ a non-negative integer) is
\begin{equation}
P_{surv}(t)=\sum_{j=-s}^s P_j(t).\label{Psurv}
\end{equation} 
We shall consider initial conditions for which $P_{surv}(0)=1$. Then, the decay of this quantity reflects how fast the walker leaves the region where it is initially located. For a classical random walk, the corresponding quantity decays as $t^{-1/2}$. For a quantum walk initialized at a single site with arbitrary coin, the survival probability decays faster than in the classical case, namely as $t^{-1}$. We will refer to this decay as the \emph{normal} decay of the quantum walk. However, if non--local initial conditions are considered, we show that the relative phase between the initial position eigenstates can be chosen so that the decay rate of the survival probability is considerably faster than the normal rate, namely $t^{-3}$. We refer to this case as an \emph{enhanced} decay.

Consider the non-local  initial conditions of the form 
\begin{equation}
\ket{\Psi^\pm}=\frac12\left(\ket{L}+i\ket{R}\right)\otimes\left(\ket{-k}\pm\ket{k} \right)
\label{init}
\end{equation} 
for a positive integer $k\le s$, so that $P_{surv}(0)=1$. 

\begin{figure}
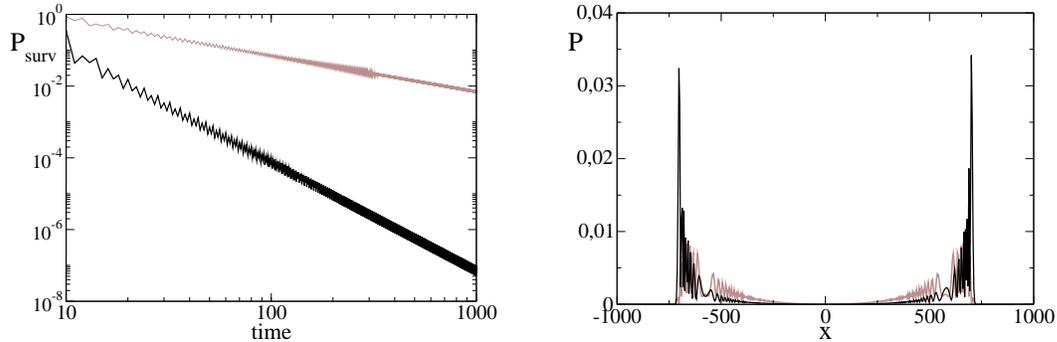

\includegraphics[width=6.5cm]{fig/ps1bw.eps}\hfill\includegraphics[width=6.5cm]{fig/dv1abw.eps}
\caption{\footnotesize Left panel: Survival probability for the quantum walk for non-local initial conditions corresponding to $k=1$ in eq.~(\ref{init}).  The different decay rates for $\ket{\Psi^-}$ (thin line)  and $\ket{\Psi^+}$ (thick line) are shown.  Right panel: Probability profile after $1000$ steps for the same initial conditions. The profile for $\ket{\Psi^-}$ (thin line) is associated to normal decay. The one corresponding to $\ket{\Psi^+}$ (thick line) shows higher peaks and corresponds to enhanced decay.  }
\label{fig:ps1}
\end{figure}

In both cases, only sites $x=\pm k$ are initially occupied, but with different phase relation. As shown in the left panel of Figure~\ref{fig:ps1}, the decay rate of the survival probability is different in both cases. For $\ket{\Psi^+}$ the survival probability decays as $t^{-3}$ (enhanced decay), while for $\ket{\Psi^-}$ the decay goes as $t^{-1}$, as for the case local initial conditions (normal decay). The probability distributions  after $1000$ steps corresponding to both initial states in eq.~(\ref{init}) are shown in the right panel of Fig.~\ref{fig:ps1}.

This behavior of the survival probability is due to an interference effect that can be understood by considering the general expression for the probability distribution, derived previously in \cite{Roma03a}. From eq.~(28) in that work, which we specialize for the case of a Hadamard walk, we have  
\begin{equation}
P_{x}(t)=\sum_{y,y'}(-1)^{(y+y')} \left[a_y(0) a_{y'}^*(0)+
b_y(0) b_{y'}^*(0)\right]\, J_{x-y}(t/\sqrt{2}) J_{x-y'}(t/\sqrt{2}),\label{dist}
\end{equation} 
where $J_\nu(\tau)$ is a cylindrical Bessel function of integer order, $\protect{\tau=t/\sqrt{2}}$ and $x,y,y'$ are integers. 
This expression, when applied to a particle initially localized at the origin, with a wavevector according to eq.~(\ref{init}) with $(k=0)$, implies that the survival probability at the origin $(s=0)$ asymptotically decays as 
$\protect{P_{surv}(t)= P_0(t)=J_0^2(t/\sqrt{2}) \sim t^{-1}}$ as $t\rightarrow\infty$, \textit{i.e.} a normal decay rate. 

We now assume an initially delocalized particle with the initial states defined in eq.~(\ref{init}). Then, eq.~(\ref{dist}) reduces to 
\begin{equation}
P_x^\pm (t)=\frac12\left[J_{x+k}(\tau)\pm  J_{x-k}(\tau) \right]^2. \label{dist-P}
\end{equation} 
In order to assess the effect of the relative phase between the position eigenstates in eq.~(\ref{init}), it suffices to consider the particular case $k=1$. 
Using the elementary recurrence relation for Bessel functions, the asymptotic (i.e. long time) form of eq.~(\ref{dist-P}) for the case $\ket{\Psi^+}$
\begin{equation}
P_x^+ (t)=\frac12\left[J_{x+1}(\tau)+  J_{x-1}(\tau) \right]^2=2x^2\left[\frac{J_x(\tau)}{\tau}\right]^2\sim \frac{1}{t^3}\qquad\mbox{for }t\gg 1
\label{dk1-sym}
\end{equation} 
is obtained. This implies an anomalous decay for this initial condition. 

On the other hand when the initial condition is $\ket{\Psi^-}$, the  probability distribution, eq.~(\ref{dist-P}), has the asymptotic form
\begin{equation}
P_x^-(t)=\frac12\left[J_{x+1}(\tau)-  J_{x-1}(\tau) \right]^2=2\left[\frac{xJ_x(\tau)}{\tau}-J_{x-1}(\tau) \right]^2\sim\frac{1}{t}\qquad\mbox{for }t\gg 1\label{dk1-antisym}
\end{equation}
and a normal decay results. 
This simple analysis can be extended to any non-local initial condition of the form (\ref{init}) with sites $\pm k$ initially occupied. The results are the same for all odd $k$. If even sites are occupied, the roles of $\ket{\Psi^\pm}$ are interchanged: for even k, $\ket{\Psi^+}$ results in normal decay and $\ket{\Psi^-}$ in enhanced decay. 

\section{Conclusions}
\label{sec:conc}
The normal asymptotic decay of the survival probability for the quantum walk on the line with localized initial conditions in position space, is $t^{-1}$. However, if non-local initial conditions with the adequate relative phases are considered, en enhanced decay rate $t^{-3}$ can be obtained. The relative phase required for enhanced decay depends on the parity of the initially loaded sites $k$ and $-k$. It is essentially an interference phenomenon and can be understood using previous analytical results for the quantum walk. As an example, we analyse in detail the case of two initial superpositions of the position eigenstates $\ket{\pm 1}$. 

On the other hand, we have recently established analytically \cite{Abal05} that the asymptotic entanglement level between the coin an position degrees of freedom, is $S_E=0.872\ldots$ for all localized initial conditions. In order to quantify entanglement, the von Neumman entropy of the reduced density matrix, $S_E$, was used. If non-local initial states are considered, any entanglement level may be obtained. In particular, for initial conditions restricted to the position subspace spanned by $\ket{\pm 1}$, the initial state $\ket{\Psi^+}$ produces the maximum level of entanglement, $S_E=0.979\ldots$ and the initial state $\ket{\Psi^-}$ produces
the minimum possible entanglement of $S_E=0.661\ldots$. Other relative phases produce intermediate entanglement levels, as described in detail in \cite{Abal05}. These results suggest that a connexion may exist between the asymptotic entanglement level and the asymptotic decay rate for the survival probability, in the sense that full asymptotic entanglement seems to be associated with the enhanced decay and in both cases non-local initial conditions are a necessary condition. Further work is required to clarify this connexion.

\vskip 1cm
\textit{G.A and R.D. acknowledge useful discussions on the subject of this work with H. Pastawski. We acknowledge support from PEDECIBA and PDT Project 29/84 (Uruguay). R.D. acknowledges financial support from  FAPERJ  (Brazil) and the Brazilian Millennium Institute  for Quantum Information--CNPq.}

\bibliography{qwnl}

\begin{thebibliography}{1}

\bibitem{Schoning}
U. Sch\"oning,
\newblock Proc. 40$^{th}$ Ann. Symp. Found. Comp. Sc., pp. 410--414, 1999.

\bibitem{Kempe03}
J. Kempe,
\newblock Contemp. Phys. 44 (2003) 307, preprint quant-ph/0303081.

\bibitem{Davidovich}
Y. Aharonov, L. Davidovich and N. Zagury,
\newblock Phys. Rev. A 48 (1993) 1687.

\bibitem{Kempe02}
J. Kempe,
\newblock Proc. of 7$^{th}$ Intern. Workshop on Randomization and Approximation
  Techniques in Comp. Sc. (RANDOM'03), pp. 354--69, 2003.

\bibitem{qw-markov}
A. Romanelli et~al.,
\newblock Physica A 338 (2004) 395, preprint quant-ph/0310171.

\bibitem{Shenvi03}
N. Shenvi, J. Kempe and B. Whaley,
\newblock Phys. Rev. A 67 (2003) 052307.

\bibitem{deco}
A. Romanelli et~al.,
\newblock Physica A 347 (2004) 137, preprint quant-ph/0403192.

\bibitem{Roma03a}
A. Romanelli et~al.,
\newblock Phys. Lett. A 313 (2003) 325, preprint quant-ph/0204135.

\bibitem{Abal05}
G. Abal et~al.,
\newblock Phys. Rev. A  (2006) to appear, preprint quant-ph/0507264.

\end{thebibliography}
\bibliographystyle{h-elsevier} 
\end{document}